# Stable nematic droplets with handles

E. Pairam[a], J. Vallamkondu[a], V. Koning[b], B. C. van Zuiden[b], P. W. Ellis[a], M. A. Bates[c],

V. Vitelli[b], A. Fernandez-Nieves[a]

[a] School of Physics, Georgia Institute of Technology, USA.

[b] Lorentz Institute for Theoretical Physics, Leiden University, The Netherlands.

[c] Department of Chemistry, University of York, UK.

## Abstract

We stabilize nematic droplets with handles against surface-tension-driven instabilities using a yield-stress material as outer fluid and study the complex nematic textures and defect structures that result from the competition between topological constraints and the elasticity of the nematic liquid crystal. We uncover a surprisingly persistent twisted configuration of the nematic director inside the droplets when tangential anchoring is established at their boundaries, which we explain after considering the influence of saddle-splay on the elastic free energy. For toroidal droplets, we find that the saddle-splay energy screens the twisting energy resulting in a spontaneous breaking of mirror symmetry; the chiral twisted state persists for aspect ratios as large as ~20. For droplets with additional handles, we observe in experiments and computer simulations that there are two additional -1 surface defects per handle; these are located in regions with local saddle geometry to minimize the nematic distortions and hence the corresponding elastic free energy.

\body

The liquid crystal in a common display is twisted due to the orientation of the molecules at the confining glass plates. By manipulating this twist using electric fields, an image can be generated. More exotic structures can emerge when the liquid crystal is confined by curved rather than flat surfaces. The topology and geometry of the bounding surface can drive the system into structures that would not be achieved without the presence of external fields. In this sense, the shape of the surface plays a role akin to an external field. Thus under confinement by curved surfaces, the molecules can self-assemble into complex hierarchical structures with emergent macroscopic properties not observed for flat liquid crystal cells. However, the design principles and properties of structures generated by this geometric route are still largely unknown.

The lowest energy state of an ordered material, such as a liquid crystal or a simple crystal, is typically defect-free since any disruption of the order will raise the elastic energy. However, the situation can be very different if the material is encapsulated within a confining volume and there is strong alignment of the molecules at the bounding surfaces. In this case, the preferred local order cannot be maintained throughout space. Such a material will be geometrically frustrated and its ground state could contain topological defects, which are spatial regions where the characteristic order of the material is lost. For nematic liquid crystals, the molecules tend to align along a common director, $\boldsymbol{n}$. The presence of defects at the boundaries, which we characterize with their topological charge, $s$, giving the amount of $\boldsymbol{n}$-rotation at the boundary as we encircle the defect, raises the energy of the system. Thus the formation of defects is normally disfavored due to this increase in energy. However, when an orientationally ordered material is confined to a closed volume, the Poincaré-Hopf theorem establishes that the total topological charge on the bounding surface must be equal to its Euler characteristic, $\chi$, a topological invariant given by $\chi = 2(1-g)$, where $g$ is the genus of the surface or its number of handles *(1)*. This theorem implies that the ground state of the system will, in many cases, incorporate topological defects. This is indeed the case when the closed surface is spherical *(2-4)*, since $\chi = 2$ for the sphere. Surfaces that are obtained by twisting, bending, stretching or generally deforming the sphere, without breaking it, are topologically equivalent, since none of these transformations introduce handles and thus they all have $\chi = 2$. In contrast, a toroidal surface is topologically different from the sphere; this has a handle and consequently $\chi = 0$.

Spherical nematics have been widely studied from experimental, theoretical and simulation points of view *(5-14)* and their intriguing technological potential for divalent nano-particle assembly has been already demonstrated *(15)*. In contrast, there are virtually no controlled experiments with ordered media in confined volumes with handles. A notable exception is the optically induced formation of cholesteric toroidal droplets inside a nematic host *(16)*. This largely reflects the difficulties in generating stable handled objects with imposed order. While the sphere is relatively easy to achieve in liquids due to surface tension, the generation of stable droplets with handles remains a formidable challenge.

In this paper, we experimentally generate stable handled droplets of a nematic liquid crystal using a continuous host with a yield stress. This approach allows us to perform the first experiments that probe nematic materials confined within droplets that are topologically different from the sphere. We observe that the toroidal nematic droplets formed are defect-free. However they exhibit an intriguing twisted structure irrespective of the aspect ratio of the torus. The stability of this configuration, which is in contrast with existent theoretical expectations *(17)*, results from the often-neglected saddle-splay contribution to the elastic free energy. Upon switching from one to multiply-handled droplets, we observe both in experiments and simulations the presence of two defects, each with topological surface charge -1, per additional handle. These defects are nucleated in regions with local saddle geometry to minimize the nematic distortions and hence the corresponding elastic free energy.

To make nematic toroidal droplets, we inject a liquid crystal (4-n-pentyl-4'-cyanobiphenyl, 5CB) through a needle into a rotating bath containing a yield-stress material consisting of (i) 1.5 wt% polyacrylamide microgels (carbopol ETD 2020), (ii) 3 wt% glycerin, (iii) 30 wt% ethanol, (iv) 1 wt% polyvinyl alcohol (PVA) and (v) 64.5 wt% ultrapure water. The presence of PVA guarantees degenerate tangential (or planar) anchoring for the liquid crystal at the surface of the droplets; we confirmed this by making spherical droplets and checking their bipolar character. We also note that the continuous phase is neutralized to pH 7, where the sample transmission is more than 90% (*18*). However, the most relevant property of this phase is its yield stress, $\sigma_y$. During formation of the torus, the stresses involved are larger than $\sigma_y$ and hence the continuous phase essentially behaves as if it were a liquid. The combination of the viscous drag exerted by the outer phase over the extruded liquid crystal and its rotational motion causes the liquid crystal to form a curved jet, as shown in Fig. 1A, which eventually closes onto itself resulting in a toroidal nematic droplet, such as that shown in Fig. 1B in bright field and in Fig. 1C between cross-polarizers. Once the torus has been formed, the elasticity of the continuous phase provides the required force to overcome the surface tension force that would naturally tend to transform the toroidal droplet into a spherical droplet. There are two ways this transformation can happen: either through a droplet break-up mechanism reminiscent of the Rayleigh-Plateau break-up of a jet into smaller spherical droplets, or through the shrinkage of the droplet towards its center to form a single spherical droplet (*19*). The relevant length scale that changes in the break-up is the tube radius, $a$, while for shrinking it is the inner radius, $R$, as defined in Fig. 1B. The minimum yield stress required to stabilize the toroidal droplet against either transformation is $\sigma_y = \gamma/a_c$ or $\sigma_y = \gamma/R_c$, where $\gamma$ is the interfacial tension between the two liquids, and $a_c$ and $R_c$ are the critical tube and inner radii of the torus below which either break-up or shrinking occurs. Using this technique we can successfully generate stable nematic toroids with an aspect ratio or slenderness $\xi = [R+a]/a$.

Remarkably, when these droplets are observed along their side view under cross-polarizers, their central region remains bright irrespective of the orientation of the droplet with respect to the incident polarization direction, as shown in Figs. 1D-F; the corresponding bright field images are shown in Figs. 1G-I. Note that for an axial torus with its director field along the tube, the cross-polarized image should appear black for an orientation of 0° and 90° with respect to the incident polarization direction. Hence our result is suggestive of a twisted structure. In fact, twisted bipolar droplets also have a central bright region, when viewed between cross-polarizers, irrespective of their orientation *(10,20-22)*. In addition, theoretical studies of DNA in toroidal geometries have also shown that the DNA condensate can be twisted as, in this case, some of the bending energy of the untwisted axial structure is released at the price of a small amount of twist energy (*17*). Interestingly, the theory predicts there is a critical value of slenderness, $\xi_c = 1.4$ for 5CB, beyond which the trade off between bend and twist energies is unfavorable and the toroidal DNA condensate remains axial.

To explore this possibility, we generate toroidal droplets with different $\xi$ and observe them between crossed-polarizers along their side view, for different orientations with respect to the incident polarization direction. We find that the central region of all droplets remains bright for all orientations, as shown in Figs. 2A and 2D for a drop with $\xi = 18.5$ and orientations of $0°$ and $45°$ with respect to the incident polarization direction. Thus, our observations suggest that we do not observe the transition from the twisted to the axial configuration predicted for toroidal DNA condensates.

To explain the lack of axial structure in our experiments, we consider the full Frank free energy:

(1) $$F = \frac{1}{2}\int dV [K_1(\nabla \cdot \boldsymbol{n})^2 + K_2(\boldsymbol{n}\cdot\nabla\times\boldsymbol{n})^2 + K_3(\boldsymbol{n}\times\nabla\times\boldsymbol{n})^2] - K_{24}\int d\boldsymbol{S}\cdot(\boldsymbol{n}\nabla\cdot\boldsymbol{n}+\boldsymbol{n}\times\nabla\times\boldsymbol{n}),$$

which, besides the well-known bulk terms representing splay, twist and bend deformations weighted with elastic constants $K_1$, $K_2$ and $K_3$, respectively, also contains the less familiar surface term representing saddle-splay deformations with elastic constant $K_{24}$. Our calculations employ an ansatz for the unit director field, $\boldsymbol{n} = n_r \boldsymbol{e}_r + n_\vartheta \boldsymbol{e}_\vartheta + n_\phi \boldsymbol{e}_\phi$, with $\boldsymbol{e}_r$, $\boldsymbol{e}_\vartheta$ and $\boldsymbol{e}_\phi$ the orthonormal basis vectors in the $r$, $\vartheta$, $\phi$ direction, respectively, and with $n_r = 0$, $n_\vartheta = \omega \dfrac{r/a}{1-\gamma r/(R+a)\cos\vartheta}$ and $n_\varphi = \sqrt{1-n_\vartheta^2}$. In these expressions, $\vartheta$ and $r$ are the polar angle and radial distance in the circular cross section of the torus, and $\phi$ is the angle in the plane perpendicular to the symmetry axis of the torus, as shown in Fig. 3A. The variational parameter $\omega$ determines the nematic organization, which continuously evolves from the axial structure, where $\omega = 0$ and hence $\boldsymbol{n} = \boldsymbol{e}_\phi$, to a twisted configuration, where $\omega \neq 0$. For simplicity, we will first set $\gamma = 1$. The resulting nematic field is then free of splay distortions and automatically obeys the tangential boundary conditions since $n_r = 0$. Moreover, detailed inspection of the nematic arrangement inside the torus reveals that the configuration is doubly twisted, as shown in Fig. 3A, where we use nails to represent the out-of-plane tilt of the director. The stable nematic organization is obtained from minimization of the elastic free energy with respect to $\omega$. After volume integration, we obtain, to leading order in $\omega$:

$$\text{(2)} \qquad \frac{F}{\pi^2 K_3 a} \approx 2\left(\xi - \sqrt{\xi^2 - 1}\right) + \frac{\left(1 - 9\xi^2 + 6\xi^4 + 6\xi\sqrt{\xi^2 - 1} - 6\xi^3\sqrt{\xi^2 - 1}\right)\xi^2 + 4(K_2 - K_{24})\xi^4 / K_3}{\left(\xi^2 - 1\right)^{3/2}} \omega^2$$

The physical implications of this equation are better seen in the limit of large $\xi$, where the Frank free energy to quartic order in $\omega$ reads:

$$\text{(3)} \qquad \frac{F}{\pi^2 K_3 a} \approx \frac{1}{\xi} + \left(4\frac{K_2 - K_{24}}{K_3}\xi - \frac{5}{4\xi}\right)\omega^2 + \frac{1}{2}\xi\omega^4$$

Note that saddle-splay acts as an external field that tends to align $\mathbf{n}$ along the $\vartheta$ direction at the surface of the torus. Similar to the Landau theory of magnetism (23), the switching of the sign of the quadratic term in Eq. 3 from positive to negative establishes a spontaneous symmetry-breaking transition from the axial to the doubly twisted configuration. Since the relevant quadratic term is zero when $\frac{K_2 - K_{24}}{K_3} = \frac{5}{16\xi_c^2}$, the relative magnitude of twist and saddle-splay determines whether the axial or the doubly twisted structure is the preferred nematic arrangement. When $\frac{K_2 - K_{24}}{K_3} > 0$, the quadratic term can be either negative or positive depending on the slenderness. Hence there is a critical $\xi_c$ above which the lowest energy state corresponds to the axial torus, as shown by the dashed line in Fig. 3B. Note, however, that $\xi_c$ can be pushed to much higher values compared to the saddle-splay free case. Below $\xi_c$, the lowest free energy state has nonzero $\omega$, corresponding to the doubly twisted torus. In this case, there are two minima of equal depth corresponding to the two possible configurations in which the handedness of the twisted nematic director is either positive or negative, as shown by the continuous line in Fig. 3B. Remarkably, when $\frac{K_2 - K_{24}}{K_3} < 0$, the quadratic term is always negative and the only possible structure is the doubly twisted configuration. This result holds irrespective of $\xi$, as shown in Fig. 3C, where we plot the phase boundary, obtained from equation 2, separating the axial from the doubly twisted regions, in a $\xi$ versus $K_{24}/K_2$ diagram. For 5CB, $K_{24} \approx K_2$ (*24-28*) and hence the axial to doubly twisted transition is either pushed to extremely slender tori or it is completely lost, consistent with our experimental observations.

We confirm our interpretation of the experimental results by first performing computer simulations of the nematic textures based on Jones calculus (*29,30*) and on the ansatz above for the director field inside the torus; these quantify how the polarization state of the incident light changes as it travels through the sample and analyzer. Consistent with the experimental results, we find that indeed the center of the torus remains bright, when viewed along its side between crossed-polarizers, irrespective of its orientation with respect to the incident polarization direction. This is shown for a nematic torus with $\xi = 2$ and $\omega = 0.4$ in Figs. 2B and 2E. Interestingly, for this aspect ratio, the center is brighter for an orientation of 0° than it is for an orientation of 45°. This is also seen experimentally (Figs. 1D and 1E). In addition, the regions to both sides of the center, encircled with a dashed line in Figs. 2B and 2E, are darker for an orientation of 0° than they are for an orientation of 45°. This is also seen experimentally (Figs. 1D and 1E). However, the situation reverses for a torus with smaller twist distortions. In this case, the center of the torus is darker at an orientation of 0° than it is for an orientation of 45°, as shown in Figs. 2C and 2F for $\omega = 0.1$. This is in agreement with the experimental results too (Figs. 2A and 2D).

We then quantify our results by measuring the twist angle in our toroidal droplets along the Z-direction, from ($r = a$, $\vartheta = 90°$) to ($r = a$, $\vartheta = 270°$), see fig. 3A. The method relies on the fact that linearly polarized light follows the twist of a nematic liquid crystal if the polarization direction is either parallel or perpendicular to the nematic director at the entrance of the sample, provided the Mauguin limit is fulfilled (*30*); the corresponding mode of propagation is referred to as extraordinary or ordinary waveguiding, respectively. We then image the torus from above (Fig. 4A), rotate the polarizer to assure that the incident polarization direction is parallel or perpendicular to the nematic director at ($r = a$, $\vartheta = 90°$) and then rotate the analyzer an angle $\phi_{exit}$ with respect to the polarizer while monitoring the transmitted intensity, $T$. The minimum in $T$, shown in Fig. 4B, reflects the lack of light propagation through the analyzer, indicating that the incident polarization direction has rotated an amount $\tau$ such that it is perpendicular to the analyzer after exiting the torus at ($r = a$, $\vartheta = 270°$). The image of the torus in this situation exhibits four black regions where extinction occurs, as shown in Fig. 4C; these corresponds to waveguiding of ordinary and extraordinary waves. It is along these regions that we measure $T$. The counterclockwise rotation of the incident polarization direction by an angle of ~60° exactly corresponds to the twist angle of the nematic along the $Z$-direction through the center of the circular cross section. However, to increase the precision of our estimate, we fit the $T$ versus $\phi_{exit}$ results to the theoretically expected transmission (*30*) leaving $\tau$ as free parameter. We find $\tau = (60.5 \pm 0.5)°$ for $\xi = 3.5$. Moreover, within the experimentally accessed $\xi$-range, we find that the twist is non-zero and that it monotonously decreases with increasing aspect ratio, as shown in Fig. 4D. Remarkably, these features are captured by our theoretical calculations for large $\xi$, as shown by the dashed line in the same figure. We note that the deviations of the experiment and the theory for small $\xi$ results from the inadequacy of the ansatz in describing the highly twisted structures observed experimentally at these low values of $\xi$. This can be partially resolved by lifting the constraint that $\gamma = 1$. This introduces a second variational parameter in the ansatz, which allows the nematic field to splay. The result qualitatively captures the experimental trend for all aspect ratios, as shown by the solid line in Fig. 4D. By further fitting the experiment to the theory in the high $\xi$-region, we obtain a value for the saddle-splay elastic constant of $K_{24} = 1.02K_2$, which is slightly larger than the twist elastic constant, confirming our previous conclusions and supporting our interpretation on the relevance of saddle-splay distortions. However, our analysis cannot exclude the possibility of a slightly smaller value of $K_{24}$ and hence of a twisted-to-axial transition for extremely large $\xi$.

Noteworthy, nematic toroids have no defects in their ground state. However, this should not be the case if we add handles since the Euler characteristic and hence the total topological charge decreases by -2 with every additional handle. However, the Poincaré-Hopf theorem provides only a conservation law that prescribes the total topological surface charge. It tells us nothing about the individual defect charges, whether they are point defects (called boojums) or singular lines inside the droplet terminating at the boundaries, the number of defects or their locations. To understand defect formation in higher genus nematic droplets, we use computer simulations of a simple nematogenic lattice model (*31*). For this model, the elastic constants are equivalent, $K_1=K_2=K_3$, and no special consideration of the saddle-splay contributions to the elastic energy is taken. Hence, we do not expect to observe any twist in the resulting structures. For the double torus, we find two types of defect configurations with comparable energy. Both of these have two defects on the surface of the double torus, each with topological charge -1. The defects are either located at the innermost regions of the inner ring of each torus, as shown in Fig. 5A, or in the outermost regions where the individual tori meet, as shown in Fig. 5B. In both cases, the defects are located in regions of local saddle geometry where the Gaussian curvature, *G*, defined as the product of the two principal curvatures, is negative. This finding shows similarity with the theoretical insight that negatively charged defects in a two-dimensional curved nematic liquid crystal are attracted to regions with negative curvature (*32-34*).

To investigate handled droplets experimentally, we exploit the elastic character of the continuous phase below the yield stress and generate two nearby single tori which are merged together by the addition of liquid crystal in the region between them. The top view of a typical droplet is shown in Fig. 5C; the corresponding cross-polarized image is shown in Fig. 5D. Interestingly, when this droplet is viewed along its side and between cross-polarizers, we observe that there is a defect in the very center of the droplet, as shown in Fig. 5E. We also observe the four-brush texture typical of $|s|=1$ defects. To determine the sign of this charge, we rotate the double torus and observe that the black brushes also rotate in the same direction (see supplementary video), indicating that the defect has a topological charge of -1 (*35*). This defect is in the back of the droplet when looked along its side, with an identical defect on its front. Hence there are two defects of charge -1 on the surface of the double torus, consistent with the constraints imposed by the Poincaré-Hopf theorem. Furthermore, they are located in regions with $G<0$, consistent with the findings of our computer simulations. We note, however, that the structure is twisted; we know this from realizing that the central region of each of the two tori forming our droplet remains bright irrespective of its side view orientation with respect to the incident polarization direction. We also note that the location of the defects obtained experimentally is consistent with just one of the two configurations obtained in the computer simulations. The other configuration has not been observed in our experiments, presumably because the way the double torus is made biases the director towards the structure in Fig. 5B.

We can also generate more complex droplets with, for example, three handles aligned along a common axis, as shown by the top view image in Fig. 5F, or arranged in a triangle, as shown in

Fig. 5G. In the first case, there are four defects, each of topological surface charge -1, located in the regions where the individual tori meet, as shown when the droplet is viewed along its side between cross-polarizers in Fig. 5H. In addition, the director is twisted, as expected based on the results for the single and double tori. In contrast, when the handles are arranged in a triangle, there are two -1 defects that cluster together in one of the three regions where the single tori meet, as indicated in Fig. 5G. In this situation, in addition to the natural frustration imposed by the bounding surface, there is an additional frustration arising from the lack of a sufficient number of negative-curvature regions between the single tori to position the defects. There are only three natural regions for the defects to be located and four defects. We find that a possible solution to this problem is to cluster two of the four defects together in one of the three natural regions for them to be located.

We have generated the first stable nematic droplets with handles using a material with a yield stress as continuous phase to stabilize these otherwise unstable droplets. Nematic toroids have no defects and exhibit a doubly-twisted configuration, similar to that observed in blue phases (*36*), irrespective of aspect ratio, which in our experiments ranges from ~2 to ~20; this results from important saddle-splay contributions to the elastic free energy. Interestingly, the comparison of the experimental measurements of the twist angle with our theoretical predictions provides a robust and simple way to measure $K_{24}$; this is important given the difficulty in determining the value of this elastic constant with current methods (*24-28*). For droplets with additional handles, we observe that there are two -1 surface defects per handle located in regions of $G < 0$ where elastic distortions are minimized.

Our work highlights the role of nematic confinement as a reliable route to induce field configurations with unique geometrical and topological properties. The chiral nematic texture observed in our toroidal droplets closely resembles a Seifert fibration of the 3-sphere, a slightly more general configuration than the celebrated Hopf fibration (39,40). Intense experimental effort has been recently directed towards constructing soft structures with non-trivial topological properties using external fields or unique sample preparation. Examples include fluid knots (41), optically created nematic torons (42), hybrid systems composed of nematic dispersions of colloidal particles with various shapes (43,38) or densely packed filamentous assemblies (44,45,17). Our experiments open up a versatile approach to generate topological soft materials that exploits nematic self-assembly within macroscopic droplets with handles, stabilized using a yield-stress material as the outer fluid.

**Acknowledgments:** We gratefully acknowledge funding from NSF (DMR-0847304), FOM and NWO. We thank Randall Kamien, Hiroshi Yokoyama, Bryan Chen and Nitin Upadhyaya for illuminating discussions.

Figures

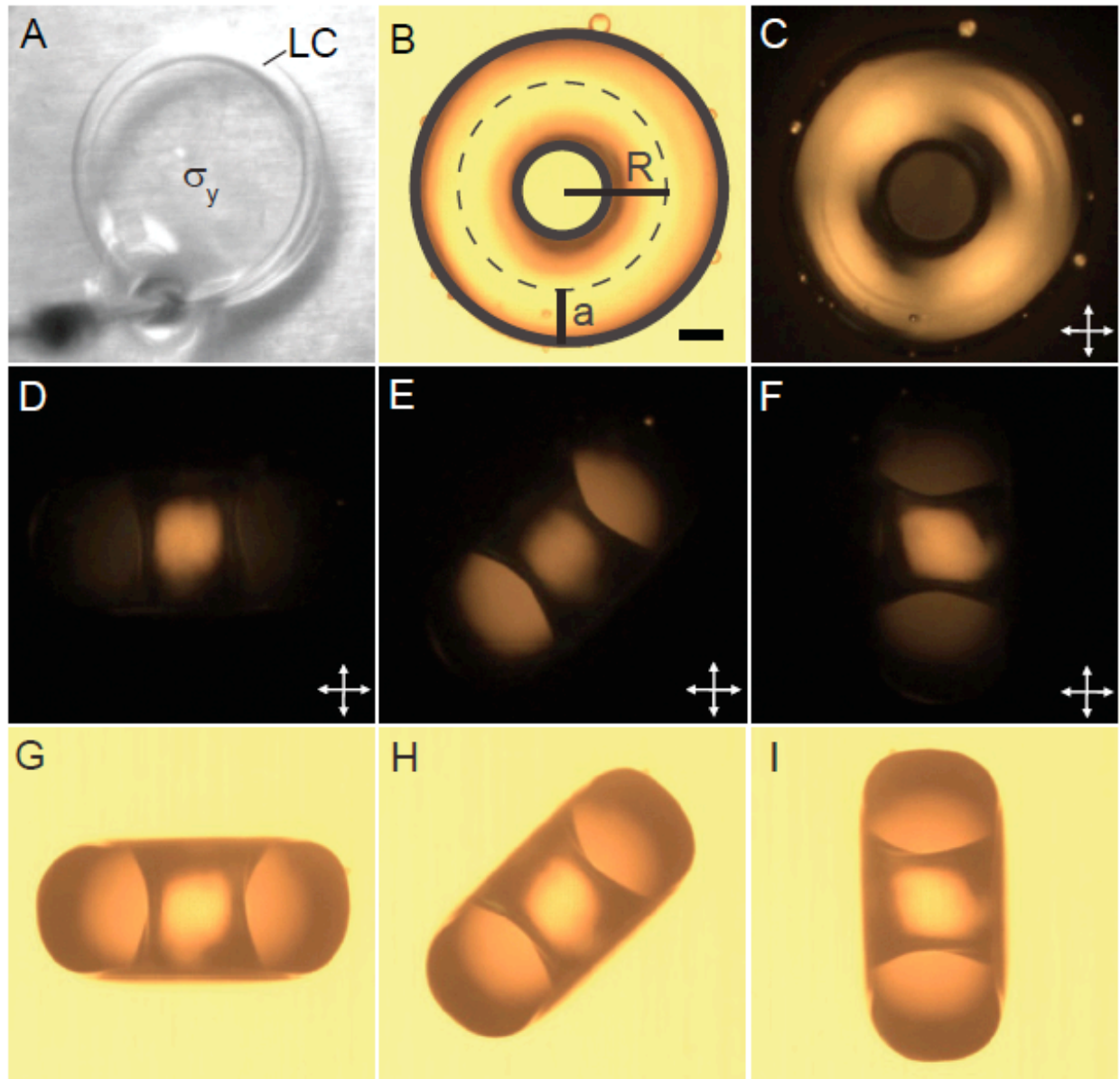

**Fig. 1. Toroidal droplets.** **(A)** Formation of a toroidal liquid crystal droplet inside a material with yield stress $\sigma_y$. The top view of a typical stable toroidal droplet of nematic liquid crystal, having tube and inner radii $a$ and $R$, is shown in **(B)** when viewed in bright field and **(C)** when viewed under cross-polarizers. **(D-F)** Side view of a typical toroidal droplet with $\xi = 1.8$ when viewed under cross-polarizers for orientations of 0º, 45º and 90º with respect to the incident

polarization direction. Note that the center part of the toroid remains bright irrespective of its orientation. **(G-I)** shows the corresponding bright field images. The dark regions of the toroid in these images are due to light refraction. Scale bar: 100 μm.

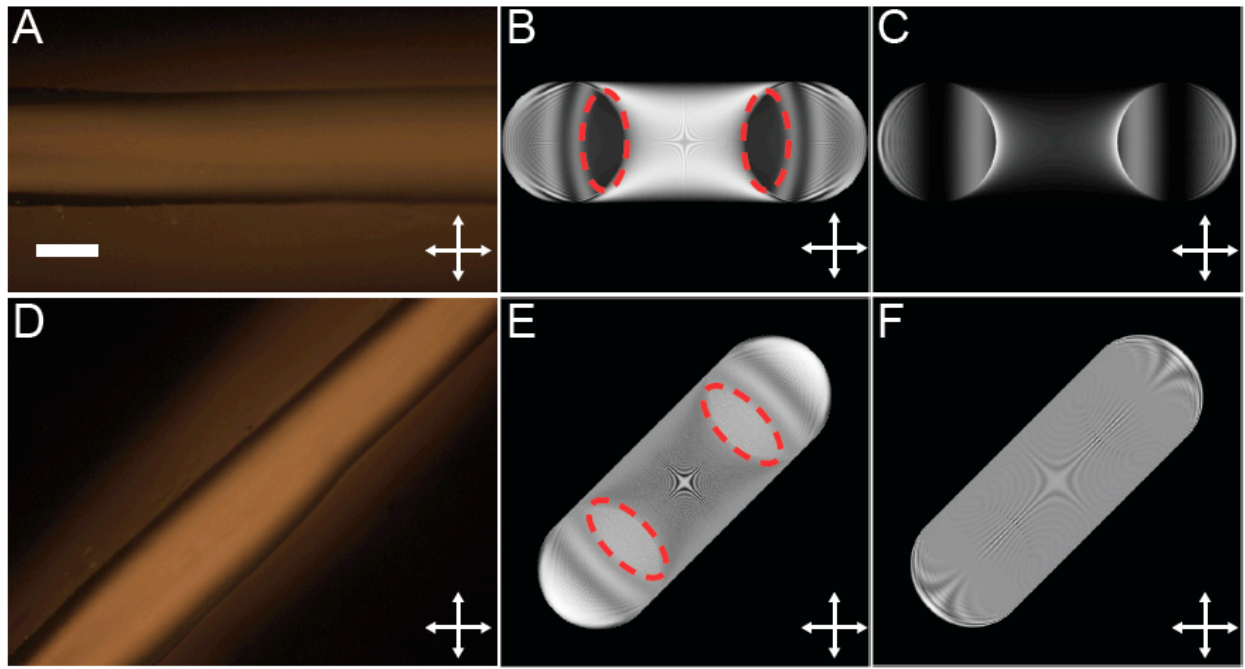

**Fig. 2: Persistence of the doubly twisted configuration with slenderness. (A,D)** Side view of a torus with $\xi = 18.5$ when viewed under cross-polarizers for orientations of 0° and 45° with respect to the incident polarization direction. Scale bar: 200 μm. **(B,C,E,F)** Computer simulation of the nematic texture of a torus with $\xi = 2$ when viewed along its side and between cross-polarizers and an orientation of 0° and 45° with respect to the incident polarization direction. In **(B,E)** $\omega = 0.4$, while in **(C,F)** $\omega = 0.1$.

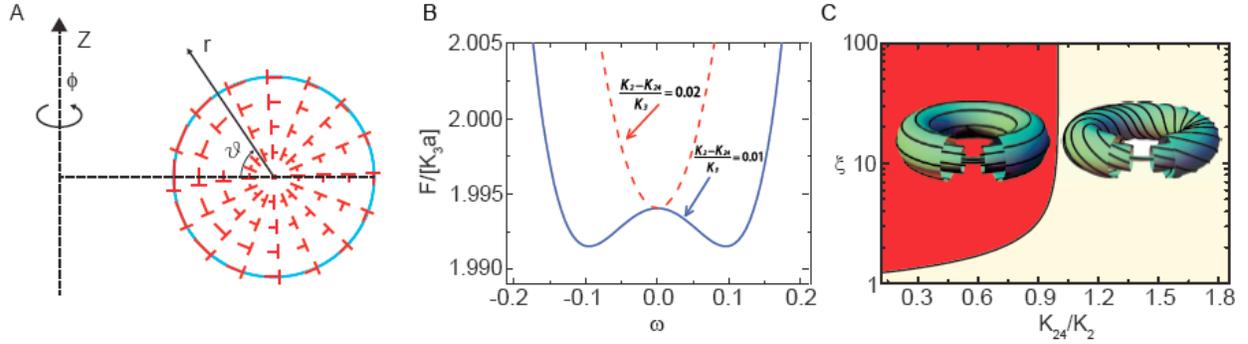

**Fig. 3. Spontaneous chiral symmetry breaking and the significance of saddle-splay distortions. (A)** Circular cross section of the torus illustrating the relevant coordinates: $\vartheta$ is the polar angle, $r$ is the radial distance from the center of the cross section and $\phi$ is the azimuthal angle. The nails indicate the tilt direction of the director; it is tilted outwards at the top, where $r = a$ and $\vartheta = 90°$, and inwards at the bottom, where $r = a$ and $\vartheta = 270°$. The presented configuration corresponds to a twisting strength $\omega = 0.49$, for a torus with aspect ratio $\xi = 2$. Note the structure is doubly twisted. The director configuration inside the whole torus is obtained by rotating the director field in this cross-section around the Z-axis. **(B)** Normalized elastic free energy, $F/(K_3 a)$, versus the variational parameter $\omega$, for $\xi=5$ and two different values of $(K_2 - K_{24})/K_3$. For $(K_2 - K_{24})/K_3 = 0.02$ (red dashed line), there is only one energy minimum at $\omega = 0$ corresponding to the axial structure shown schematically on the left in panel **(C)**. For $(K_2 - K_{24})/K_3 = 0.01$ (blue continuous line), there are two minima located at $\omega \approx \pm 0.1$ corresponding to the two possible handedness of the doubly twisted structure shown schematically on the right in panel **(C)**. The ratio $K_{24}/K_2$ determines whether there is a transition between the axial and the doubly twisted structure and if so what is the critical value of $\xi$, or whether the doubly twisted structure remains irrespective of $\xi$. This is shown in the structural phase diagram of **(C)**, where we have used that $K_2 = 0.3 K_3$ for 5CB (*35*). Since for 5CB,

$K_{24} \approx K_2$ (24-28) the axial to double twist transition is either completely lost or shifted to very high values of $\xi$, consistent with our experimental observations.

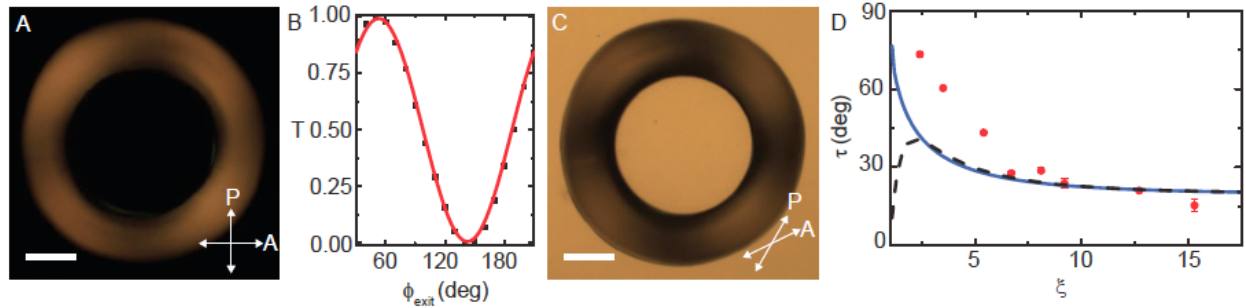

**Fig. 4. Determination of the twist angle and its dependence with slenderness. (A)** Top view of a torus with $\xi = 3.5$ when viewed from the top and between cross-polarizers. **(B)** Transmission, $T$, as a function of the angle between the incident polarization direction and the analyzer, $\phi_{exit}$. The line is a fit to the theoretical expectation in the Mauguin limit (30) with the twist angle, $\tau$, as the only free parameter. We obtain $\tau = (60.5 \pm 0.5)°$. **(C)** Top view of the same torus at the minimum of the transmission curve. We measure T along the four black regions that are observed, which are darkest for the indicated direction of the polarizer and the analyzer. **(D)** Twist angle as a function of $\xi$. The dashed line represents the theoretical prediction based on equation (2), for $K_{24} = 1.02\ K_2$. The solid line represents the theoretical prediction based on the improved ansatz including the second variational parameter $\gamma$ for the same value of $K_{24}$, where we have used that $K_1 = 0.64\ K_3$ for 5CB (35). Scale bar: 200 µm.

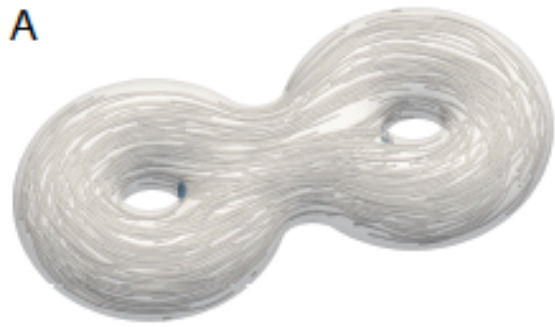 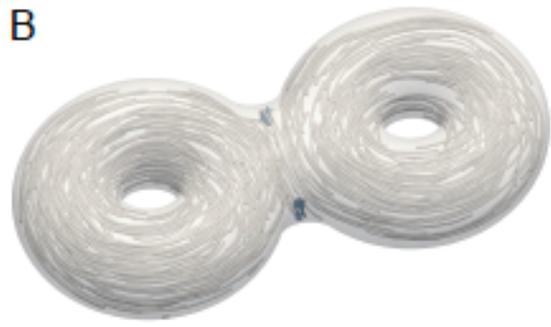
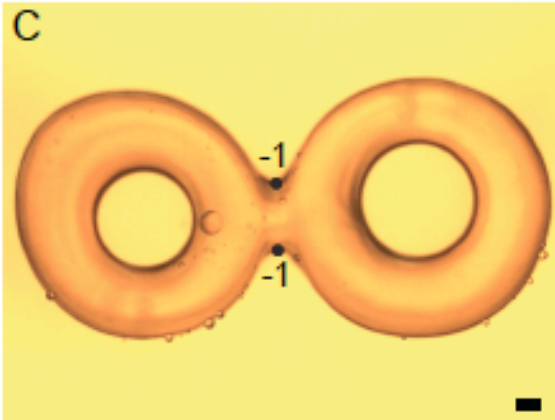 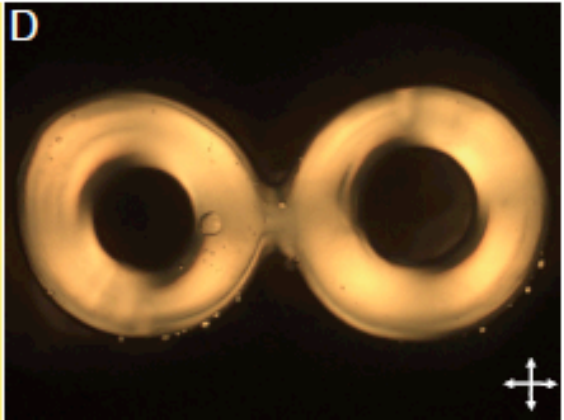
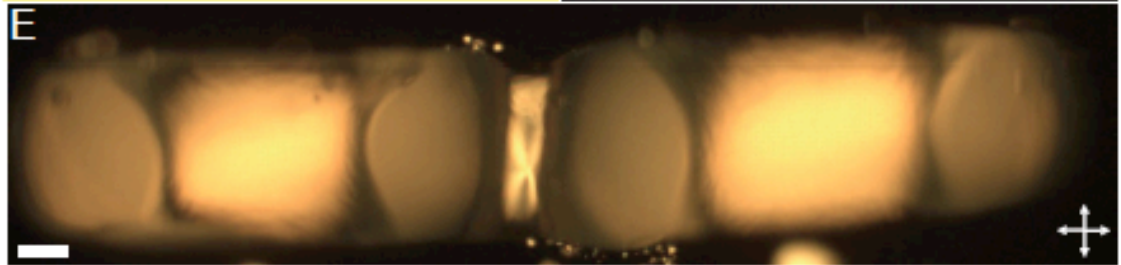
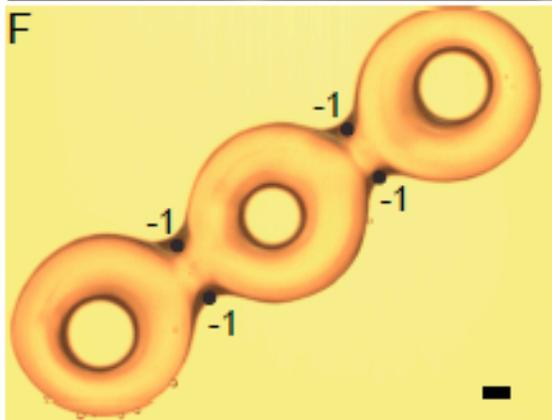 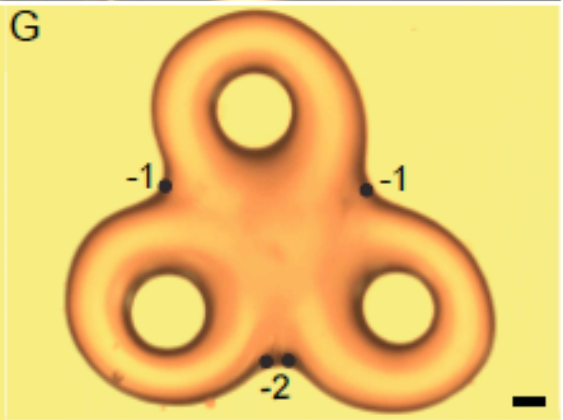
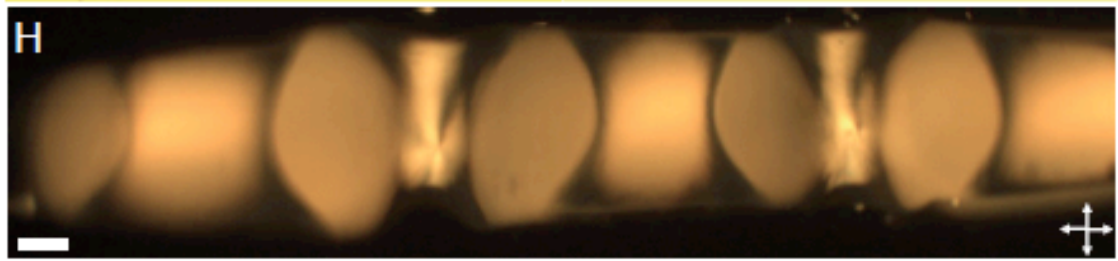

**Fig. 5. Double and triple toroids. (A,B)** The two textures found by computer simulations of a typical double torus. They both have two defects (solid spots) on the surface of the double torus, each with topological charge -1, located in regions of negative Gaussian curvature, either **(A)** at the innermost regions of the inner ring of each torus or **(B)** at the outermost regions where the individual tori meet. **(C-H)** Experimental double and triple toroids: **(C)** Top view of a double toroid in bright field. Solid dark circles indicate the location of the two -1 surface defects. **(D)** The same image under cross-polarizers. **(E)** Side view of the double toroid under cross-polarizers when focused at its back. The four black brushes in the region where the two single toroids meet indicate the presence of a topological defect with charge $|s|=1$. The sign of this charge is determined by rotating the double torus. Since the brushes rotate in the same sense as the rotation, we conclude the defect has charge $s=-1$ (see supplementary video). By changing the focal plane, we confirm there is another $s=-1$ defect at the front of the double toroid. **(F)** Top view image of a triple toroid with a side-by-side arrangement of the handles. **(G)** Top view image of a triple toroid with a triangular arrangement of the handles. Solid circles show the defect locations found by looking at the droplets between cross-polarizers along different viewing directions. **(H)** Side view of a triple toroid with a side-by-side arrangement of the handles viewed under cross-polarizers. The defects are located in the outer regions where the individual toroids meet. Scale bar: 100 μm.